\documentstyle[11pt]{article}
\begin{document}
\newcommand{\BE}{\begin{equation}}
\newcommand{\EE}{\end{equation}}
\newcommand{\BA}{\begin{eqnarray}}
\newcommand{\EA}{\end{eqnarray}}
\newcommand{\de}{{\rm d}}
\vspace*{25mm}
\begin{center}
{\LARGE\bf Stochastic Quantization of Autonomous ${\bf \Phi^4}$}
\vspace{20mm}\\
{\Large U. Ritschel}
\vspace{13mm}\\
{\large\it
Fachbereich Physik, Universit\"at GH Essen, D-45117 Essen
(F R Germany)}
\vspace{43mm}\\
\end{center}
{\bf Abstract:} The non-perturbative autonomous renormalization
of the scalar $\Phi^4$-model
is applied in the framework of stochastic quantization.
I show that this requires
a selective, momentum-dependent renormalization
of the Onsager coefficient $\lambda$, a direct consequence
of the characteristic wavefunction renormalization applied.
As a result, I obtain a Langevin equation for the renormalized
constant mode of the field, which is solved numerically.
It is demonstrated for temperature zero
that, starting from specified initial conditions, the system relaxes
to its equilibrium state, the symmetry-breaking vacuum of the
``static''
$\Phi^4$-theory.
\newpage
Recently there has been much interest in the autonomous
renormalization (AR) of the Higgs field \cite{auto}.
The AR emerges very naturally
as an alternative solution to the renormalization group equation
\cite{cons,fisi}, where the
major difference to the conventional perturbative procedure
is an infinite wavefunction renormalization
\cite{cost,latt}.
Concerning the number of free parameters the AR turns out to be very
economical; there is only one for the scale invariant theory .
As a consequence, it is possible to make a concrete prediction for
the mass of the Higgs boson,
which comes out surprisingly heavy, at about 2 TeV
\cite{cons,rodr,nghi}. For more technical and phenomenological
details concerning the AR,  I refer to
the literature cited above.

The question posed in this Letter is whether the AR can also be
applied
in the context of stochastic quantization. The basic idea
of this method is to assign a purely relaxational dynamics in a {\it
fictitious}
time coordinate, which I will simply call time
in the following,
to the system \cite{stoc}. Starting from some non-equilibrium initial
state, on account of ergodicity
and detailed balance, the system relaxes to its equilibrium state,
the ground state of the euclidean quantum field theory. The latter
will be called the {\it static} theory below.
The dynamics is described by a Langevin equation
and has many formal analogues in statistical physics. There, however,
a real time-dependent process, like the dynamics of a
ferromagnet close to its Curie point, is described by the equations.
Reasons for introducing stochastic quantization have been advantages
in gauge theories \cite{stoc} and numerical simulations \cite{schu}.

Consider now the scale invariant euclidean
$\Phi^4$ theory in a finite periodic
hypercube with volume $\Omega=L^d$, described by the action
\BE
S[\phi_B] = \int_{\Omega} {\rm d}^dx \left[\frac{1}{2} ({\bf\nabla}
\phi_B)^2
+ \frac{g}{4!}\, \phi_B^4\right]\;.
\EE
{}From the static calculation
one knows that the AR requires a {\it selective}
wavefunction renormalization, {\it i.e.}, one
has to distinguish between constant and finite-momentum (FM)
modes \cite{cost}.
While the FM modes are essentially
not renormalized,
the (bare) constant mode has to be rescaled by a logarithmically
divergent $Z$-factor.
In a previous work on the static theory \cite{fisi}, it was
demonstrated that the most natural starting point for the AR is a
finite system. Although one is not really interested in finite-size
effects,
many features of the model, like the convexity of the effective
potential, the emergence of spontaneous symmetry breaking, and
finite-temperature effects, can best
be understood by starting from a finite system and then taking
carefully the infinite-volume limit. Much the same strategy will be
pursued
in this Letter for the stochastically quantized model. For the sake of
simplicity, the calculations will be restricted to the
zero-temperature
case. To treat the more general situation, one
dimension of the volume $\Omega$
has to be kept finite \cite{fisi}.

Following Consoli and Stevenson \cite{cost}, I denote (in slightly
adapted
form)
the relation between
bare and renormalized Fourier amplitudes of the field by
\BE\label{ware}
\tilde\phi_B({\bf p},t)=Z({\bf p})^{\frac12}\,\tilde\phi({\bf p},t)=
\left[\sqrt{3}\,\epsilon^{-\frac12}\,\delta_{{\bf p},{\bf 0}}
+(1-\delta_{{\bf p},{\bf 0}})\right]\,\tilde\phi({\bf p},t)\>.
\EE
Since I will use dimensional regularization in the following,
logarithmic divergences are expressed by powers of $1/\epsilon$
(with $\epsilon=4-d$) from
the start \cite{dime}.  The vector ${\bf p}$ in
(\ref{ware}) has $d$ components with
discrete values $p_i=2\pi k_i/L$, where $k_i\in
Z\!\!\!\!\hspace{.2mm}Z$. The factor $\sqrt{3}$ in (\ref{ware}) is a
result
of the static calculation \cite{cons}. It comes from a
finite rescaling
of the constant mode
(called $z_0$ in \cite{cons}),
which guarantees the proper normalization of the (static) propagator.

The dynamics assigned to the
system is described by the Langevin equation
\BE
\partial_t\, \tilde\phi_B({\bf p},t) +\lambda_B({\bf p})
\frac{\delta S[\phi_B]}{\delta\tilde \phi_B({\bf -p},t)}
=\tilde \xi_B({\bf p},t) \,, \label{langevin}
\EE
where $\tilde\xi_B({\bf p},t)$ is a
Gaussian random ``force'' with mean zero and variance
\BE\label{vari}
\langle\tilde \xi({\bf p},t)\,\tilde\xi(B{\bf q},t')\rangle =
2\,\lambda_B({\bf p})\,\delta_{{\bf p},-{\bf q}}\,\delta(t-t') \;.
\EE
{}From work on classical statistical dynamics \cite{jans},
it is well known
that the renormalization of the Onsager coefficient $\lambda$ is
closely
related to the wavefunction renormalization. Hence, from this
point of view, it is natural to infer a relation
of the form
\BE\label{lamb}
\lambda_B({\bf p})=\left[z_{\lambda}\,\epsilon^{\sigma}\,
\delta_{{\bf p},{\bf 0}}
+(1-\delta_{{\bf p},{\bf 0}})\right] \lambda\>,
\EE
where the finite normalization constant $z_{\lambda}$ and the exponent
$\sigma$ are to be determined below.

Like in the static calculation \cite{fisi}, the first step
towards a solution of (\ref{langevin}) is the
approximative integration of
the FM-modes in a one-loop type procedure. Similar methods have
been applied in the context of critical dynamics in finite-size
systems
\cite{diehl,gold}.
Here I am following closely the method suggested by Goldschmidt
\cite{gold}.
After integrating out the FM modes,
one obtains the Langevin equation for the constant mode:
\BE\label{laco}
\partial_t\,\phi_0(t)+\epsilon^{\sigma}z_{\lambda}\lambda\>\frac{g}{6}
\left(3\,C(t)+\phi_0(t)^2\right)\phi_0(t)=\xi_0(t)\,,
\EE
where I have defined
$$
\phi_0(t)\equiv\Omega^{-\frac12}\,\tilde \phi_B({\bf 0},t)
\qquad{\rm such\, \>that}\qquad \phi_0(t) =\frac{1}{\Omega}\int{\rm
d}^dx\>\phi_B({\bf x},t)\,,
$$
and $\xi_0(t)$ is related to $\tilde\xi({\bf 0},t)$ analogously.

In (\ref{laco}), the term
$C(t)$ represents the
effect of the FM modes. It is
the time-dependent analogue of the tad-pole graph of the
static one-loop
calculation \cite{fisi} and may be represented in the form
\BE\label{sum1}
C(t)=2\,\lambda\,L^{-d}\left.\int_0^{t}\,{\rm d }s
\sum_{{\bf p}\neq {\bf 0}}\exp\left( -2\,\lambda\,\int_{t-s}^t{\rm
d}s'\,
\left[{\bf p}^2+\frac{g}{2}\,\phi_0^2(s')\right]\right)
\right.\>.
\EE
Suppose that for large time $\phi_0(t)$ approaches a
constant equilibrium value. Then (\ref{sum1})
reduces to the well-known static result
\BE\label{sum2}
C(t=\infty)=L^{-d}\left.\sum_{{\bf p}\neq {\bf 0}}\frac{1}{{\bf
p}^2+g\,\phi_0(t=\infty)^2/2}\right. \,.
\EE
For the further evaluation,
(\ref{sum1}) may be rewritten as
\BE\label{cvte}
C(t)=2\lambda\,L^{-d}\int_0^t{\rm d}s\>{\cal
B}\!\left(\frac{8\pi^2\lambda s}{L^2}\right)\>
\exp\left(-\lambda \,g\int_0^{s}{\rm d}s'\,\phi_0(t-s')^2\right)
\EE
with
$$
{\cal B}(x)={\cal A}(x)^d-1\qquad{\rm and}\qquad {\cal
A}(x)=\sum_{k=-\infty}^{\infty}\,
{\rm e}^{- k^2 \,x} \>.
$$
The fact that the sum in (\ref{sum1}) does not extend over
zero momentum has no influence on the UV-behavior.
Consequently, $C(t)$
contains the typical
$1/\epsilon$ pole. In (\ref{cvte}) it is caused
by the divergence of the integrand for small
argument,  where ${\cal B}(x)\sim (\pi/x)^{d/2}$.

At this point a brief
comment on the {\it initial conditions} is appropriate\,: For the
evolution of the field expectation, one initial condition will be to
assign some specified value to the constant mode at $t=0$. Moreover,
from
(\ref{sum1}) and (\ref{cvte}), it becomes obvious that another
``initial condition''
is $C(t=0)=0$, {\it i.e.}, fluctuations of the FM modes are suppressed
at $t=0$, which, in a sense, emerges here as the natural
initial state of the relaxational process.

For the further evaluation of (\ref{cvte}) it is convenient to isolate
the pole by rewriting the whole expression in the form
\BE\label{rewr}
C(t) = C_{\rm div}(t)+\Delta C(t) + {\cal O}\left(L^{-d}\right).
\EE
The contributions ${\cal O}\left(L^{-d}\right)$ do not influence the
dynamics in the infinite-volume limit.

The UV-divergent part $C_{\rm div}$ is given by
\BE\label{cdiv}
C_{\rm div}(t)=\frac{(2\lambda)^{1-d/2}}{(4\pi)^{d/2}}\int_0^t{\rm
d}s\,s^{-d/2}
\exp\left(-\lambda\,g\,s\,\phi_0(t)^2\right)\,.
\EE
The integral can be calculated exactly, and, after
analytic continuation to $4-\epsilon$ dimensions, the result
may be expanded
in a power series in $\epsilon$\,:
\BE\label{cres}
C_{\rm div}(t) = -\frac{g\phi_0(t)^2}{(4\pi)^2\epsilon} +
C^{(0)}_{\rm div}(t)+{\cal O}(\epsilon)\,,
\EE
where the finite contribution is given by
\BE\label{cfin}
C^{(0)}_{\rm div} =
-\frac{{\rm e}^{-\Theta}}{(4\pi)^2 \,2\lambda t}+
\left.\frac{g\,\phi_0^2}{2(4\pi)^2}\right[\left.
\log\left(\frac{g\,\phi_0^2}{16 \pi^2\, \Phi_v^2}\right)
-\log\left(\Theta\right){\rm e}^{-\Theta}
-\left.\frac{{\rm d}\,\gamma\left(x,\Theta\right)}{{\rm d}x}
\right|_{x=1}-C_E\right]
\EE
with the abbreviation
$\Theta=\Theta(t)=\lambda t\,g \phi_0^2(t)$.
In (\ref{cfin}), $\gamma(x,y)$ denotes the incomplete Gamma function
and
$C_E$ Euler's constant.
$\Phi_v$ emerges in this stage of the calculation as an arbitrary
momentum scale
generated by dimensional transmutation, but its normalization
is chosen such that it later on also turns out as
the vacuum expectation value of the field.

The second term on the rhs of (\ref{rewr}) is a memory term given by
\BE\label{memo}
\Delta C(t)= \frac{(2\lambda)^{1-d/2}}{(4\pi)^{d/2}}
\int_0^t\,{\rm d}s\, s^{-d/2}\left[\exp\left(-\lambda g\int_{t-s}^{t}
{\rm d}s'\,\phi_0(s')^2\right)-\exp\left(-\lambda
s\,g\phi_0(t)^2\right)\right]\>.
\EE

Like in the static calculation \cite{fisi}, the pole in
(\ref{cdiv}) is cancelled by the nonlinear term in (\ref{laco})
by assigning the UV-flows
\BE\label{uvflow}
\phi_0(t) = \sqrt{3}\,\epsilon^{-1/2} \Phi(t)
\qquad{\rm and}\qquad
g=\frac{(4\pi)^2\epsilon}{3}
\EE
to the constant mode ({\it cf.} Eqn.\,(\ref{ware})) and the
bare coupling, respectively. After this cancellation, the remainder
$C^{(0)}_{\rm div}(t)+\Delta C(t)$ on the lhs of (\ref{laco})
is of order $\epsilon^0$.
Thus, together with
the factor $g$, the deterministic ``force'' would be of order
$\epsilon$.
A finite contribution may be obtained by choosing $\sigma = -1$
in (\ref{lamb}). Moreover, the normalization $z_{\lambda}$ can be
fixed
by demanding that the deterministic part is equal to the derivative
of the {\it static} effective potential for $t \rightarrow \infty$.
In this limit, the memory term $\Delta C(t)$ tends to zero, and only
the first term in square brackets in (\ref{cfin}) survives.
As a result, one derives the relation
\BE\label{deri}
\frac{\partial V(\Phi)}{\partial \Phi}= \frac{4 \pi^2 z_{\lambda}}{3}
\Phi^3\log\left(\frac{\Phi^2}{\Phi_v^2}\right)\>,
\EE
from which it is also obvious that $\Phi=\Phi_v$ (where the logarithm
has its zero)
corresponds to
the asymmetric ground
state.
Now the second
derivative of $V$ with respect to the renormalized field
must be equal to the squared propagator mass, called
$M_H$ in the following. The latter can
be read from (\ref{sum2}), and
inserting the UV-flows (\ref{uvflow}) one finds
$$M_H^2=8\pi^2\,\Phi_v^2\,.$$ Eventually,
the second derivative of $V$, calculated from (\ref{deri}), is equal
to
$M_H^2$ if $z_{\lambda}=3$.

What, then, happens to the noise in view of finite and infinite
renormalizations
of the field? Carrying out the rescalings
in (\ref{laco}) gives rise to a renormalized Langevin equation of the
form
$\partial_t \Phi(t) + {\rm finite}=3^{-1/2}\epsilon^{1/2} \xi(t)$.
Thus, the most natural UV-flow for the {\it constant} mode of the
noise is
\BE
\xi(t)=\sqrt{3}\,\epsilon^{-1/2}\,\Xi(t)\>.
\EE
The renormalized noise $\Xi(t)$ has mean zero, and with (\ref{vari})
and (\ref{lamb}) one obtains for the variance\,:
\BE
\langle \Xi(t)\Xi(t')\rangle = 2\lambda\,L^{-d}\, \delta(t-t')\, .
\EE
The average magnitude of the noise is ${\cal O}(L^{-d/2})$,
and, therefore, in the infinite-volume limit the dynamics becomes
essentially
deterministic. This is completely analogous to the static case, where
a
saddle-point approximation for the path integral over the constant
mode
becomes exact in the infinite-volume limit \cite{fisi}.

With (\ref{cfin}), (\ref{memo}), and (\ref{uvflow}) in
(\ref{laco}) and after introducing dimensionless variables
$\bar t= 2\,\lambda\, t \,M_H^2$ and $\bar\Phi={\Phi}/{\Phi_v}$, the
equation of motion for the constant mode reads
\BA\label{fieq}
\partial_{\bar t}\,\bar\Phi&=&\frac14\,\bar\Phi
\,\left\{\bar\Phi^2\left[
\log(\bar t\,\bar\Phi^2)\exp(-\bar t\,\bar\Phi^2)+\left.\frac{{\rm
d}\,\gamma\left(x,\bar t\,\bar\Phi^2\right)}{{\rm d}x}
\right|_{x=1}+C_E-\log(\bar\Phi^2)\right]\right.\nonumber\\
& &\left.+\frac{\exp(-\bar t\,\bar\Phi^2)}{\bar t}- \int_0^{\bar
t}\,{\rm d}
s\, s^{-2}\left[\exp\left(-\int_{\bar t-s}^{\bar t}{\rm
d}s'\,\bar\Phi(s')^2
\right)-\exp\left(-s\,\bar\Phi^2\right)\right] \right\}\, ,
\EA
where the argument of $\bar\Phi$ has been suppressed when it is $\bar
t$.

There is one special feature, which one encounters on the
way to a numerical solution of (\ref{fieq}).
The first term on the rhs of (\ref{cfin}) and, in turn, the first
term in the second line of (\ref{fieq}) are singular for
$\bar t\rightarrow 0$. This is due to the initial state discussed
above,
and the singularity could be avoided by choosing different, though
more complicated, initial conditions.
Similar so-called
short-time singularities are well known from statistical
physics, where they give rise to anomalous behavior of relaxational
processes in their early stages \cite{insli}. In stochastic
quantization,
where one is mainly interested in the equilibrium state, the
short-time
behavior seems to be merely a technical problem.
A non-singular equation may be obtained with the ansatz $\bar\Phi(\bar
t)
=\bar t^{\,1/4}\Psi(\bar t\,)$ with finite initial value $\Psi(0)$.

A number of numerical solutions is depicted in the Figure.
$\Psi(0)$ serves as a parameter.
As a consequence of the short-time singularity, all trajectories of
$\bar\Phi$ have to start from zero. Then the expectation value
increases
and overshoots the equilibrium value,
but asymptotically, for large time, all solutions approach
$\bar\Phi=1$ from above.
There are no solutions
that cross the origin.

In summary, I have applied the autonomous renormalization
to the stochastically quantized $\Phi^4$-theory.
The most important
difference compared with conventional perturbation theory is a
selective
renormalization of the Onsager coefficient $\lambda$ as formulated in
Eqn.\,(\ref{lamb}). A finite equation of motion, which becomes purely
deterministic in the infinite-volume limit, has been derived and
solved
numerically. Asymptotically, all solutions
approach the equilibrium state, the
asymmetric ground state of the static $\Phi^4$-model.

\newpage\noindent
{\Large\bf Figure captions}\vspace{1cm}\\
{\bf Fig.} Numerical solutions of Eqn.\,(\ref{fieq}) with the ansatz
$\bar\Phi(\bar t)=\bar t^{\,1/4}\Psi(\bar t)$, where $\Psi(0)$ serves
as
the parameter.
 \end{document}